# Second-order Dirac equation of graphene electrons in an electromagnetic field and their novel spin

Ji Luo[a]

*Department of Physics and Institute for Functional Nanomaterials,*

*University of Puerto Rico at Mayagüez, Mayagüez, PR 00681, USA*

The second-order Dirac equation (DE) and its velocity operator of graphene electrons in an electromagnetic field are obtained according to tight-binding $\vec{k} \cdot \vec{p}$ method. With extra terms included, they demonstrate graphene's electronic motion more completely through a more complete Ehrenfest theorem and present finer properties of graphene electrons. Eigen-energy given by the second-order DE for field-free graphene indicates that extra terms may affect the trembling motion of graphene electrons. For graphene in a magnetic field, eigen-energy given by the second-order DE suggests that graphene electrons have a new kind of spin of a boson other than true electronic spin and the Dirac particle's pseudo-spin, which will modify graphene's properties such as its optical spectra.

---

[a] Email address: ji.luo@upr.edu

## I. INTRODUCTION

The theoretical importance of graphene originates from Dirac equation (DE) of its electronic states, and interesting properties have been predicted according to this equation.[1,2] DE is the basis of electronic states of graphene both field-free[3] and in an external field.[4-6] It is also the basis of graphene's quantum Hall effect.[7-9] Recently electronic wave packets of graphene attracted research interest, especially for graphene in a magnetic field, and time-dependent (TD) DE was applied in the investigations.[10-15] Because of DE, graphene electrons are often treated by analogy with Dirac particles and graphene is regarded as a useful material for verifying quantum electrodynamics results.[1,2] Since DE describes particles with spin, relations between spin and graphene lattice have also been studied.[16]

While DE is widely applied to graphene investigations, its derivation is not so extensively discussed in the literature as its application.[17] In particular, for graphene in an electromagnetic field, DE is obtained simply by adding potentials to that for field-free graphene, called the minimal coupling or Peierls substitution.[2,18] Nevertheless, graphene's experiments have employed greater and greater magnetic fields.[19] The validity of DE for great fields is an interesting issue and can be established by examining its derivation. More importantly, the derivation may reveal extra terms that have been neglected and their finer effects on graphene properties. Besides, although wave packets play a crucial role in graphene's transport properties, investigations of their general propagation in an electromagnetic field are still

lacking.[20]

In this work, the derivation of TD DE for graphene electrons in an orthogonal electromagnetic field by the tight-binding (TB) $\vec{k}\cdot\vec{p}$ method is examined.[21] This derivation leads to the second-order DE and its velocity operator which contain extra terms. The position, momentum, velocity, and angular momentum of a Dirac particle described by this second-order DE are obtained from those of an electron described by original Schrödinger equation. Time-derivatives of these quantities are investigated and a more complete Ehrenfest theorem is derived according to the second-order DE. These results are used to investigate the general motion of graphene electrons. The same TB $\vec{k}\cdot\vec{p}$ method is also used to investigate the role of graphene's lattice potential in the electronic motion. Solutions of the second-order DE for field-free graphene and for graphene in a pure magnetic field are obtained and found to reveal new finer properties of graphene electrons, especially their novel spin in the magnetic field. Significance of the results to quantum mechanics is discussed.

## II. DERIVATION OF THE SECOND-ORDER DIRAC EQUATION

The graphene is taken as the $xy$-plane with $x$-axis parallel to one set of carbon-carbon bonds. Unit vectors of the axes are denoted by $\vec{x}$, $\vec{y}$, and $\vec{z}$. Throughout this work, $\tau = \pm 1$ are respectively for two kinds of Dirac points that correspond to Fermi wave vectors $\vec{k}_F = (2\pi/3\sqrt{3}a_0)(\sqrt{3}\vec{x} + \tau\vec{y})$ with $a_0 = 0.142$ nm the carbon-carbon bond length. Wave-vector difference $\vec{k} - \vec{k}_F$ is denoted by $\vec{k}$ for simplicity. Besides, combined subscripts are used such as $\psi_{1,2}$ for $\psi_1$ and $\psi_2$ respectively, etc..

Starting from the single-electron Hamiltonian of field-free graphene

$$\hat{H}_0 = -\frac{\hbar^2}{2m}\nabla^2 + V \tag{1}$$

with $m$ the mass of the electron and $V$ the lattice potential including the average effect of other electrons, TB $\vec{k}\cdot\vec{p}$ method expresses an electronic eigen-state near a Dirac point by[21]

$$\psi_{\vec{k}}(\vec{r}) = e^{i\vec{k}\cdot\vec{r}}[C_1(\vec{k})\psi_A(\vec{r}) + C_2(\vec{k})\psi_B(\vec{r})], \tag{2}$$

$$\psi_{A,B}(\vec{r}) = \Omega^{1/2} e^{\pm\pi i/12} \sum_{A,B} e^{i\vec{k}_F\cdot\vec{R}_{A,B}} \phi(\vec{r}-\vec{R}_{A,B}), \tag{3}$$

where $\psi_{A,B}$ are Bloch functions at $\vec{k}_F$ respectively for two kinds of atoms $A$ and $B$, $\vec{R}_{A,B}$ are the position vectors of the atoms, $\phi(\vec{r})$ is the perpendicular $p_z$ orbital of carbon atoms, and $\Omega$ is the area of a unit cell. The role of factors $e^{\pm\pi i/12}$ will be discussed later. For graphene in a parallel electric field $\vec{E} = -\nabla\varphi$ and a perpendicular magnetic field $\vec{B} = \nabla\times\vec{A}$ with $\varphi(\vec{r})$ and $\vec{A}(\vec{r})$ the scalar and vector potentials respectively, the single-electron Hamiltonian becomes

$$\hat{H} = \frac{1}{2m}(-i\hbar\nabla + e\vec{A})^2 + V - e\varphi. \tag{4}$$

The electromagnetic field is regarded as perturbation and an electronic eigen-state is expressed as the superposition of nonperturbated states (2), or

$$\psi_\varepsilon(\vec{r}) = \int_\infty C_\varepsilon(\vec{k})\psi_{\vec{k}}(\vec{r})d^2\vec{k}, \tag{5}$$

where $\varepsilon$ is the eigen-energy and $C_\varepsilon$ is the superposition coefficient. Further, a time-dependent state such as a wave packet is the superposition of different

eigen-states (5), or

$$\psi(\vec{r},t) = \int_\infty C(\varepsilon)\psi_\varepsilon(\vec{r})e^{-i\varepsilon t/\hbar}d\varepsilon, \qquad (6)$$

with $C(\varepsilon)$ the superposition coefficient. According to Eqs. (2) and (5) $\psi$ can be expressed as[21]

$$\psi(\vec{r},t) = \psi_1(\vec{r},t)\psi_A(\vec{r}) + \psi_2(\vec{r},t)\psi_B(\vec{r}), \qquad (7)$$

where $\psi_{1,2}$ given by

$$\psi_{1,2}(\vec{r},t) = \int_\infty d\varepsilon \int_\infty C(\varepsilon)C_\varepsilon(\vec{k})C_{1,2}(\vec{k})e^{i\vec{k}\cdot\vec{r}-i\varepsilon t/\hbar}d^2\vec{k} \qquad (8)$$

are two TD envelope functions that vary gently at atomic scale.

Wave function $\psi$ in Eq. (7) satisfies TD Schrödinger equation

$$i\hbar\frac{\partial \psi}{\partial t} = \hat{H}\psi \qquad (9)$$

with $\hat{H}$ given by Eq. (4). To obtain $\psi_{1,2}$ form Eq. (9), a set of approximations for the integrals of $\phi$ is applied. First, as $\phi$ is a real function localized at an atom, the only nonzero overlapping integrals are those for the same atoms, that is,[22]

$$\int_\infty \phi^2(\vec{r}-\vec{R}_{A,B})d^3\vec{r} = 1. \qquad (10)$$

Second, the only nonzero momentum integrals are those for two nearest atoms, that is,[23]

$$\int_\infty \phi(\vec{r}-\vec{R}_{A,B})\nabla\phi(\vec{r}-\vec{R}_{B,A})d^3\vec{r} = \frac{2}{3\hbar}mv_F\vec{n}_{AB,BA}, \qquad (11)$$

where $v_F \approx 10^6$ ms$^{-1}$ is the Fermi velocity, and $\vec{n}_{AB,BA}$ are unit vectors along $\overrightarrow{AB}$ or $\overrightarrow{BA}$ respectively. Accordingly, the sum of the three neighboring momentum integrals centered at an atom $A$ or $B$ is respectively

$$\sum_{B,A} e^{i\vec{k}_F \cdot (\vec{R}_{B,A} - \vec{R}_{A,B})} \int_\infty \phi(\vec{r} - \vec{R}_{A,B})(-i\hbar\nabla)\phi(\vec{r} - \vec{R}_{B,A}) d^3\vec{r} = mv_F e^{\pm\pi i/6}(\vec{x} \mp \tau i\vec{y}), \quad (12)$$

where in application, factors $e^{\pm\pi i/6}$ will be canceled by $e^{\pm\pi i/12}$ in $\psi_{A,B}$. Finally, a gentle function $f(\vec{r},t)$ that multiplies $\phi(\vec{r} - \vec{R}_{A,B})$ or $\nabla\phi(\vec{r} - \vec{R}_{A,B})$ is regarded as fixed at $f(\vec{R}_{A,B},t)$ in the integration area and thus can be moved out from the integrals. Besides, $\phi$ and $\psi$ depend on coordinate $z$, but $\psi_{1,2}$, $\varphi$, and $\vec{A}$ do not. Their integrals are triple or double ones respectively.

By substituting Eq. (7) into Eq. (9) one obtains

$$i\hbar\left(\psi_A \frac{\partial \psi_1}{\partial t} + \psi_B \frac{\partial \psi_2}{\partial t}\right)$$
$$= \frac{1}{2m}[\psi_A(-i\hbar\nabla + e\vec{A})^2\psi_1 + \psi_B(-i\hbar\nabla + e\vec{A})^2\psi_2] - e\varphi(\psi_A\psi_1 + \psi_B\psi_2)$$
$$- \frac{\hbar^2}{m}(\nabla\psi_A \cdot \nabla\psi_1 + \nabla\psi_B \cdot \nabla\psi_2) - \frac{i\hbar e}{m}(\psi_1 \vec{A} \cdot \nabla\psi_A + \psi_2 \vec{A} \cdot \nabla\psi_B) \quad (13)$$
$$- \frac{\hbar^2}{2m}(\psi_1 \nabla^2 \psi_A + \psi_2 \nabla^2 \psi_B) + V(\psi_1\psi_A + \psi_2\psi_B).$$

The last two brackets vanish because $\psi_{A,B}$ are eigen-states of Hamiltonian (1) at $\vec{k}_F$ where the eigen-energy is set to zero and thus $-(\hbar^2/2m)\nabla^2\psi_{A,B} + V\psi_{A,B} = 0$. One chooses an atom $A$ or $B$ respectively, multiplies Eq. (13) by $\phi(\vec{r} - \vec{R}_{A,B})$ and conducts the integration. According to the integral approximations of $\phi$ one obtains

$$\begin{cases} i\hbar \dfrac{\partial \psi_1}{\partial t} = \dfrac{1}{2m}(-i\hbar\nabla + e\vec{A})^2 \psi_1 - e\varphi\psi_1 + v_F(\vec{x} - \tau i\vec{y}) \cdot (-i\hbar\nabla + e\vec{A})\psi_2 \\ i\hbar \dfrac{\partial \psi_2}{\partial t} = \dfrac{1}{2m}(-i\hbar\nabla + e\vec{A})^2 \psi_2 - e\varphi\psi_2 + v_F(\vec{x} + \tau i\vec{y}) \cdot (-i\hbar\nabla + e\vec{A})\psi_1 \end{cases}, \quad (14)$$

where all functions take values at the chosen $\vec{R}_{A,B}$. Equation (14) is the second-order TD DE that contains second-order terms. The second-order Hamiltonian (SH) that operates on the two-component wave function $\Psi = (\psi_1 \ \psi_2)^T$ is

$$\hat{H}_s = \left(\frac{1}{2m}\hat{\vec{p}}^2 - e\varphi\right)I_2 + v_F\vec{\sigma}\cdot\hat{\vec{p}}, \tag{15}$$

where $\hat{\vec{p}} = -i\hbar\nabla + e\vec{A}$ is the momentum operator, $I_2$ is the $2\times 2$ unite matrix, and $\vec{\sigma} = \sigma_x\vec{x} + \tau\sigma_y\vec{y}$ with $\sigma_{x,y}$ the first two Pauli matrixes. Only if the second-order term $(1/2m)\hat{\vec{p}}^2$ is neglected, SH (15) reduces to usual Dirac-like Hamiltonian (DH)

$$\hat{H}_D = v_F\vec{\sigma}\cdot\hat{\vec{p}} - e\varphi I_2. \tag{16}$$

## III. PHYSICAL QUANTITIES BASED ON THE SECOND-ORDER DIRAC EQUATION

Originally, graphene electrons are described by $\psi$ in Eq. (7) and the two-component wave function $\Psi = (\psi_1\ \psi_2)^T$ describe virtual Dirac particles or the so-called Dirac Fermions. In principle expectation values of quantum mechanics operators should be calculated from $\psi$ in Eq. (7). By using the method of deriving Eq. (14), the results can be expressed in terms of $\Psi = (\psi_1\ \psi_2)^T$ and one thus obtains expectation values for the Dirac particle. Properties of the electron thus can be presented by those of the Dirac particle.

For a gentle function $f(\vec{r},t)$, according to Eq. (3) and the integral approximations of $\phi$ one has

$$\int_\infty f(\vec{r},t)\psi^*_{A,B}\psi_{A,B}d^3\vec{r} = \Omega\sum_{A,B}f(\vec{R}_{A,B},t) = \int_\infty f(\vec{r},t)d^2\vec{r},$$

$$\int_\infty f(\vec{r},t)\psi^*_A\psi_B d^3\vec{r} = \int_\infty f(\vec{r},t)\psi^*_B\psi_A d^3\vec{r} = 0,$$

$$\int_\infty f(\vec{r},t)\psi^*_{A,B}(-i\hbar)\nabla\psi_{B,A}d^3\vec{r} = mv_F(\vec{x}\mp\tau i\vec{y})\Omega\sum_{A,B}f(\vec{R}_{A,B},t) = mv_F(\vec{x}\mp\tau i\vec{y})\int_\infty f(\vec{r},t)d^2\vec{r}$$

, $\int_\infty f(\vec{r},t)\psi^*_{A,B}(-i\hbar)\nabla\psi_{A,B}d^3\vec{r}=0$. By applying these results to $\psi$ in Eq. (7) one obtains

$$\int_\infty f(\vec{r},t)\psi^*\psi d^3\vec{r} = \int_\infty f(\vec{r},t)\Psi^+\Psi d^2\vec{r}, \tag{17}$$

$$\int_\infty f(\vec{r},t)\psi^*\hat{\vec{p}}\psi d^3\vec{r} = \int_\infty f(\vec{r},t)\Psi^+\hat{\vec{p}}\Psi d^2\vec{r} + mv_F\int_\infty f(\vec{r},t)\Psi^+\vec{\sigma}\Psi d^2\vec{r} \tag{18}$$

with $\Psi^+ = (\psi_1^* \ \psi_2^*)$.

The cell-averaged probability density of the electron is $\rho = (1/\Omega)\int_{\Omega\times\infty}\psi^*\psi d^3\vec{r}$. According to the integral approximations of $\phi$, it becomes $\rho = |\psi_1(\vec{R}_A)|^2 + |\psi_2(\vec{R}_B)|^2$, or

$$\rho(\vec{r},t) = \Psi^+\Psi. \tag{19}$$

The $\psi_{A,B}$ given by Eq. (3) contain the area instead of the number of the unit cell and thus guarantee normalization $\int_\infty \Psi^+\Psi d^2\vec{r} = 1$ if $\int_\infty \psi^*\psi d^3\vec{r} = 1$. The expectation value of the position operator $\hat{\vec{r}} = \vec{r}$ for the electron is $\vec{r} = \int_\infty \psi^*\vec{r}\psi d^3\vec{r}$. According to Eq. (17) one obtains

$$\vec{r} = \int_\infty \Psi^+\vec{r}\Psi d^2\vec{r}. \tag{20}$$

The electron and the Dirac particle thus have the same position, and wave packets $\psi$ and $\Psi$ describe the same quasi-particle motion. The expectation value of the momentum of the electron is $\vec{p}_e = \int_\infty \psi^*\hat{\vec{p}}\psi d^3\vec{r}$. According to Eq. (18), one obtains

$$\vec{p}_e = \int_\infty \Psi^+\hat{\vec{p}}\Psi d^2\vec{r} + mv_F\int_\infty \Psi^+\vec{\sigma}\Psi d^2\vec{r}. \tag{21}$$

The expectation value of the momentum operator of the Dirac particle is

$$\vec{p}_D = \int_\infty \Psi^+\hat{\vec{p}}\Psi d^2\vec{r}. \tag{22}$$

The electron and the Dirac particle thus have different momentums and

$$\vec{p}_e = \vec{p}_D + m v_F \int_\infty \Psi^+ \vec{\sigma} \Psi d^2 \vec{r} . \tag{23}$$

The expectation value of the velocity operator $\hat{\vec{v}}_e = (1/m)\hat{\vec{p}}$ for the electron is $\vec{v} = \int_\infty \psi^* \hat{\vec{v}}_e \psi d^3 \vec{r} = (1/m)\vec{p}_e$. According to Eq. (21), one has

$$\vec{v} = \frac{1}{m} \int_\infty \Psi^+ \hat{\vec{p}} \Psi d^2 \vec{r} + v_F \int_\infty \Psi^+ \vec{\sigma} \Psi d^2 \vec{r} . \tag{24}$$

The electron and the Dirac particle should have the same velocity, since they have the same position. Equation (24) indicates that the velocity operator and the current density of the Dirac particle are respectively

$$\hat{\vec{v}}_D = \frac{1}{m} I_2 \hat{\vec{p}} + v_F \vec{\sigma} , \tag{25}$$

$$\vec{j}(\vec{r},t) = \frac{1}{m} \Psi^+ \hat{\vec{p}} \Psi + v_F \Psi^+ \vec{\sigma} \Psi . \tag{26}$$

Only if $(1/m)\int_\infty \Psi^+ \hat{\vec{p}} \Psi d^2 \vec{r}$ is neglected, one obtains $\vec{v} = v_F \int_\infty \Psi^+ \vec{\sigma} \Psi d^2 \vec{r}$ and the velocity operator $v_F \vec{\sigma}$ corresponding to DH (16). The expectation value of the angular momentum operator $\hat{\vec{l}} = \hat{\vec{r}} \times \hat{\vec{p}}$ for the electron is $\vec{l}_e = \int_\infty \psi^* \hat{\vec{l}} \psi d^3 \vec{r}$. According to Eq. (18), one obtains

$$\vec{l}_e = \int_\infty \Psi^+ \hat{\vec{l}} \Psi d^2 \vec{r} + m v_F \int_\infty \Psi^+ \vec{r} \times \vec{\sigma} \Psi d^2 \vec{r} . \tag{27}$$

The expectation value of the angular momentum of the Dirac particle is

$$\vec{l}_D = \int_\infty \Psi^+ \hat{\vec{l}} \Psi d^2 \vec{r} . \tag{28}$$

The electron and the Dirac particle thus have different angular momentums and

$$\vec{l}_e = \vec{l}_D + m v_F \int_\infty \Psi^+ \vec{r} \times \vec{\sigma} \Psi d^2 \vec{r} . \tag{29}$$

**IV. EHRENFEST THEOREM OF THE SECOND-ORDER DIRAC EQUATION**

One then calculates $d\vec{r}/dt$ starting from $\vec{r} = \int_\infty \psi^*\vec{r}\psi d^3\vec{r}$ and $\vec{r} = \int_\infty \Psi^+\vec{r}\Psi d^2\vec{r}$ respectively. Both Shrödinger equation (9) and the second-order DE (14) lead to

$$\frac{d\vec{r}}{dt} = \vec{v}. \tag{30}$$

By applying the second-order DE (14) to $\vec{p}_D = \int_\infty \Psi^+\hat{p}\Psi d^2\vec{r}$ and $\int_\infty \Psi^+\vec{\sigma}\Psi d^2\vec{r}$ one obtains

$$\frac{d\vec{p}_D}{dt} = -e\int_\infty (\rho\vec{E} + \vec{j}\times\vec{B})d^2\vec{r}, \tag{31}$$

$$\frac{d}{dt}\int_\infty \Psi^+\vec{\sigma}\Psi d^2\vec{r} = -\vec{z}\times\frac{2\tau v_F}{\hbar}\int_\infty \Psi^+\sigma_z\hat{\vec{p}}\Psi d^2\vec{r}, \tag{32}$$

with $\sigma_z$ the third Pauli matrix. According to Eqs. (23) and (24) one has

$$\frac{d\vec{p}_e}{dt} = -e\int_\infty (\rho\vec{E} + \vec{j}\times\vec{B})d^2\vec{r} - \hat{z}\times\frac{2\tau m v_F^2}{\hbar}\int_\infty \Psi^+\sigma_z\hat{\vec{p}}\Psi d^2\vec{r}, \tag{33}$$

$$\frac{d\vec{v}}{dt} = -\frac{e}{m}\int_\infty (\rho\vec{E} + \vec{j}\times\vec{B})d^2\vec{r} - \vec{z}\times\frac{2\tau v_F^2}{\hbar}\int_\infty \Psi^+\sigma_z\hat{\vec{p}}\Psi d^2\vec{r}. \tag{34}$$

On the other hand, one may calculate $d\vec{p}_e/dt$ starting from $\vec{p}_e = \int_\infty \psi^*\hat{\vec{p}}\psi d^3\vec{r}$. According to Eq. (9), one obtains[24,25]

$$\frac{d\vec{p}_e}{dt} = -e\int_\infty \psi^*\vec{E}\psi d^3\vec{r} - \frac{e}{m}\int_\infty \psi^*\hat{\vec{p}}\psi\times\vec{B}d^3\vec{r} - \int_\infty \psi^*\nabla V\psi d^3\vec{r}. \tag{35}$$

According to Eqs. (17) and (18), the first two integrals of Eq. (35) become $-e\int_\infty (\rho\vec{E} + \vec{j}\times\vec{B})d^2\vec{r}$. By comparing Eq. (35) with Eq. (33), one has

$$-\int_\infty \psi^*\nabla V\psi d^3\vec{r} = -\vec{z}\times\frac{2\tau m v_F^2}{\hbar}\int_\hbar \Psi^+\sigma_z\hat{\vec{p}}\Psi d^2\vec{r}. \tag{36}$$

Equation (17) does not apply to $\int_\infty \psi^*\nabla V\psi d^3\vec{r}$, because lattice force $-\nabla V$ is not a gentle function at atomic scale. Equation (36) thus demonstrates the relation between lattice force and the Dirac particle's wave function. The two integrals in Eq. (34) respectively represent Lorentz force of the external field and lattice force. By applying Eq. (14) to $\vec{l}_D = \int_\infty \Psi^+ \hat{\vec{l}} \Psi d^2\vec{r}$ and $\int_\infty \Psi^+ \vec{r} \times \vec{\sigma} \Psi d^2\vec{r}$ one obtains

$$\frac{d\vec{l}_D}{dt} = -e\int_\infty \vec{r} \times (\rho\vec{E} + \vec{j} \times \vec{B})d^2\vec{r}, \tag{37}$$

$$\frac{d}{dt}\int_\infty \Psi^+ \vec{r} \times \vec{\sigma} \Psi d^2\vec{r} = -\hat{z}\frac{2\tau v_F}{\hbar}\int_\infty \Psi^+ \sigma_z (\vec{r} \cdot \hat{\vec{p}} - i\hbar)\Psi d^2\vec{r}. \tag{38}$$

According to Eq. (29) one has

$$\frac{d\vec{l}_e}{dt} = -e\int_\infty \vec{r} \times (\rho\vec{E} + \vec{j} \times \vec{B})d^2\vec{r} - \hat{z}\frac{2\tau m v_F^2}{\hbar}\int_\infty \Psi^+ \sigma_z (\vec{r} \cdot \hat{\vec{p}} - i\hbar)\Psi d^2\vec{r}. \tag{39}$$

By calculating $d\vec{l}_e/dt$ from $\vec{l}_e = \int_\infty \psi^* \hat{\vec{l}} \psi d^3\vec{r}$ and compare the result with Eq. (39) one has

$$-\int_\infty \vec{r} \times \psi^* \nabla V \psi d^3\vec{r} = -\hat{z}\frac{2\tau m v_F^2}{\hbar}\int_\infty \Psi^+ \sigma_z (\vec{r} \cdot \hat{\vec{p}} - i\hbar)\Psi d^2\vec{r}. \tag{40}$$

Equation (40) demonstrates the relation between the momentum of lattice force and the Dirac particle's wave function.

Equations (30) and (34) constitute Ehrenfest theorem that determines quasi-particle motion of graphene electrons in the electromagnetic field.[24,25] The velocity of the wave packet is determinative to graphene's electronic transport such as trembling motion and quantum Hall effect.[2,26] Fine difference will be resulted between the velocity determined by the second-order DE (14) and that by usual DE with DH (16). As a result, the second-order DE (14) should be necessary if, for instance, the trembling motion of graphene electrons is investigated.[2] On the other hand, one may use approximations of deriving the second-order DE (14) to estimate lattice force $-\int_\infty \psi^* \nabla V \psi d^3 \vec{r}$. $\nabla V$ has the same $C_3$ symmetry as the two-dimensional graphene crystal, with the axis parallel to $\vec{z}$ and passing through an atom. Besides, $\phi$ is axially symmetric with respect to the same axis. As a result, for an atom $A$ or $B$, one has $\int_\infty \phi^2(\vec{r}-\vec{R}_{A,B}) \nabla V d^3\vec{r} = 0$. $\nabla V$ also has mirror symmetry with respect to the graphene plane, the perpendicular plane containing a carbon-carbon bond, and the perpendicular bisector plane of a carbon-carbon bond. Therefore for two neighboring atoms $A$ and $B$, one has $\int_\infty \phi(\vec{r}-\vec{R}_A) \phi(\vec{r}-\vec{R}_B) \nabla V d^3\vec{r} = 0$. According to Eq. (7) and integral approximation $\phi$, one obtains

$$-\int_\infty \psi^* \nabla V \psi d^3\vec{r} = 0. \tag{41}$$

Equations (36) and (32) respectively lead to

$$\int_\infty \Psi^+ \sigma_z \hat{\vec{p}} \Psi d^2\vec{r} = 0, \tag{42}$$

$$\int_\infty \Psi^+ \vec{\sigma} \Psi d^2\vec{r} = \text{const.} \tag{43}$$

As a result, the lattice force in Ehrenfest theorem (34) vanishes and the electronic motion is determined by the Lorentz force. Usually the field applied to graphene is gentler than the wave packet, especially as a uniform one. In this case $\vec{E}$ and $\vec{B}$ can be moved out from the integrals of Eq. (34) and one obtains Newtown equation

$$\frac{d\vec{v}}{dt} = -\frac{e}{m}(\vec{E} + \vec{v} \times \vec{B}), \tag{44}$$

where $\vec{E}$ and $\vec{B}$ take their values at $\vec{r} = \int_\infty \Psi^+ \vec{r} \Psi d^2\vec{r}$. Equation (44) is a necessary condition and real electronic motion still depends on eigen-states of SH (15) and their superposition. In fact, according to DH (16), the velocity of graphene electrons in the electromagnetic field does not vary with time.[27] This amounts to Eq. (43) of SH. In the case of the second-order DE, the dependence of the velocity on time is approximately given by Eq. (44). More exact result should be obtained according to Eq. (34).

## V. TWO SOLUTIONS TO THE SECOND-ORDER DIRAC EQUATION

Eigen-states of SH are more difficult to obtain than those of DH except for some special cases.[27,28] As an example, for field-free graphene with $\vec{E}=0$, $\vec{B}=0$, eigen-energies and eigen-states of SH (15) are

$$\varepsilon = \frac{\hbar^2 k^2}{2m} \pm v_F \hbar k ,\qquad(45)$$

$$\begin{pmatrix}\psi_1\\\psi_2\end{pmatrix} = \begin{pmatrix}k\\\pm(k_x+\tau i k_y)\end{pmatrix} e^{i\vec{k}\cdot\vec{r}},\qquad(46)$$

with $\vec{k} = k_x \vec{x} + k_y \vec{y}$ the wave vector. Although eigen-states (46) are the same as those of the DH (16), the small term $\hbar^2 k^2 / 2m$ of the eigen-energy will modify the fine motion of graphene electrons. According to Eq. (6), the same superposition coefficient $C(\varepsilon)$ leads to different wave packets for SH (15) and DH (16). As a result, the trembling motion of electrons will be different if the second-order DE is used.

Another example is graphene in a pure uniform magnetic field with $\vec{E}=0$, $\vec{B}=B\vec{z}$. Eigen-energies and eigen-states of SH (15) are

$$\varepsilon = \frac{n\hbar eB}{m} \pm \sqrt{2n v_F^2 \hbar eB + \frac{\hbar^2 e^2 B^2}{4m^2}} ,\qquad(47)$$

$$\begin{pmatrix} \psi_1 \\ \psi_2 \end{pmatrix} = \begin{pmatrix} [\varepsilon - (2n+1)\hbar eB/2m]H_{n-1}(\xi) \\ iv_F \sqrt{\hbar eB} H_n(\xi) \end{pmatrix} e^{-\xi^2/2 + ik_y y}, \tag{48}$$

where $n = 0, 1, 2, \cdots$, $\xi = \sqrt{eB/\hbar}\, x + \sqrt{\hbar/eB}\, k_y$, $H_n(\xi) = (-1)^n e^{\xi^2} d^n e^{-\xi^2}/d\xi^n$ is the $n$th order Hermite polynomial, and $H_{-1}(\xi) = 0$. Equation (48) gives eigen-states for $\tau = 1$. For $\tau = -1$, eigen-states can be obtained by exchanging $\psi_1$ and $\psi_2$.

For any magnetic field now available, one has $\hbar eB/m \ll v_F \sqrt{\hbar eB}$.[18] According to the approximation $\sqrt{1+x} \approx 1 + x/2$ for a small $x$, eigen-energy (47) becomes $\varepsilon \approx n\hbar eB/m \pm v_F \sqrt{2n\hbar eB} \pm \hbar^2 e^2 B^2 / 8m^2 v_F \sqrt{2n\hbar eB}$. The last term can be neglected and one obtains

$$\varepsilon \approx \frac{n\hbar eB}{m} \pm v_F \sqrt{2n\hbar eB}. \tag{49}$$

In the right-hand side of Eq. (49), the first term can be expressed as $2n\mu_B B$ with $\mu_B = \hbar e/2m$ the Bohr magneton, and the second term is the Landau level (LL) given by DH (16). The two-component wave functions obtained from DE describe pseudo-spin of Dirac particles of graphene and the true spin of graphene electrons are usually not considered in most investigations. Eigen-energy (49) indicates that graphene electrons in a magnetic field can be regarded as having a new kind of spin

$$s_z = n\hbar \tag{50}$$

whose magnitude, for an exited LL, is greater than the true spin $\hbar/2$. As a result, if effects of graphene electrons' true spin such as Zeeman splitting are concerned, those of the new spin (50) should also be considered. Spin (50) is that of a boson. Graphene electrons in a magnetic field may thus resemble bosons in some aspects.

## VI. CONCLUDING REMARKS

In conclusion, the second-order DE is necessary if finer properties of graphene electrons are to be obtained. The novel spin of graphene electrons resulting from the second-order DE is worth further investigation and may probably lead to their boson properties. More interesting results are possible if eigen-states of SH (15) for a more general electromagnetic field can be obtained. The second-order DE enriches the relation between the Dirac particle's wave function and its classical motion through a more complete Ehrenfest theorem, where both Lorentz force of the external field and graphene's lattice force are explicitly presented. Since lattice force vanishes, graphene may act as an ideal material for investigating classical motion of its electrons in the external field. Apart from device application, this may provide insight into the relation between quantum mechanics and classical one, since wave functions can be obtained through the second-order DE. Graphene thus may also be helpful to the understanding of some fundamental problems of quantum mechanics.